\documentstyle[preprint,tighten,aps,epsfig]{revtex}

\newcommand{\gcu}{\mbox{$\, \stackrel{ > }{ _{\sim} } \,$}}
\newcommand{\lcu}{\mbox{$\, \stackrel{ < }{ _{\sim} } \,$}}

\begin{document}

\author{S. Degl'Innocenti $^{1,2}$, P.G. Prada Moroni$^{1,2}$, 
B. Ricci$^{3,4}$\\
  $^1$ Dipartimento di Fisica, Universit\`a di Pisa,
       Largo B. Pontecorvo 3,  56126 Pisa, Italy\\
  $^2$ INFN Sezione di Pisa, Largo B. Pontecorvo 3,  56126 Pisa, Italy\\
  $^{3}$Istituto Nazionale di Fisica Nucleare, Sezione di Ferrara,
        via Paradiso 12, I-44100 Ferrara, Italy, \\
  $^{4}$Dipartimento di Fisica dell'Universit\`a di Ferrara,
        via Paradiso 12, I-44100 Ferrara, Italy
}

\title{The heavy elements mixture and the stellar cluster age}

\maketitle

\begin{abstract}
Recent analyses of solar spectroscopic data (see Asplund, Grevesse \& Sauval
2004 and references therein) suggest a significative variation of the heavy
elements abundance. The change of heavy mixture might affect the
determination of the clusters age for two different reasons: the change of
theoretical isochrones at fixed metallicity and the variation of the inferred
cluster metallicity from the observed [Fe/H].  The first point is analyzed 
discussing the effects of updating the metal distribution on theoretical
evolutionary tracks and isochrones for metallicities suitable for the galactic
globular and open clusters and for the bulk population of the Large Magellanic
Clouds.  The maximum variation of the estimated age ($\approx$0.5 Gyr),
although not negligible, is still within the present uncertainty.  The
second point is addressed by comparing present theoretical predictions with
the very precise observational data of the Hyades cluster from the Hypparcos
satellite.
\end{abstract}

\section{Introduction}

Through the years the knowledges in stellar physics have been continuously
improved thanks to the progress in the determination of the physical inputs
for the stellar models (nuclear reaction rates, equation of state, opacity
coefficients, microscopic diffusion etc..) and in the observational
capabilities. Now the general scenario is well defined and confirmed by an
huge amount of observational data of the Sun and of the field and cluster
stars in our Galaxy. However several problems are still not completely solved
(e.g. the precise treatment of external convection, the overshooting and
diffusion efficiency, the discrepancy between theory and observation for the
light elements surface abundance etc..). 

Recent analyses of spectroscopic data using three dimensional hydrodynamic
atmospheric models (see Asplund et al. 2004 and references therein) have
reduced the derived abundances of CNO and other heavy elements with respect to
previous estimates (Grevesse \& Sauval 1998, hereafter GS98).  Thus the Z/X
solar value decreases from the GS98 value (Z/X)$_{\odot}$=0.0230 to
(Z/X)$_{\odot}$=0.0165. GS98 already improved the mixture by Grevesse \& Noels
(1993, hereafter GN93), widely adopted in the literature, mainly revising the
CNO and Ne abundance and confirming the very good agreement between the new
photospheric and meteoric results for iron.

Several works analyzed the effects of the last update of the composition on
the solar characteristics pointing out a disagreement between the observed and
the predicted sound speed (see e.g. Bahcall et al. 2005, Basu \& Antia 2004).
However, further investigations are needed because the still present
uncertainties on the physical inputs adopted in the models prevent the
reaching of a firm conclusion.

The change of heavy mixture might affect also the determination of the
clusters age for two different reasons: 1) the change of theoretical
isochrones at fixed metallicity; 2) the variation of the inferred cluster
metallicity from the observed [Fe/H].

 In the first part of the paper we will analyse the former point, while
the last section is devoted to the comparison with Hipparcos observational
data for the Hyades cluster. We point out that our aim is not to generally
discuss the effects of a variation of the heavy elements abundance, as already
done by several authors for Population II stars (see e.g. Simoda \& Iben
1968,1970, Renzini 1977, Castellani \& Tornamb\`e 1977, Rood 1981, Rood \&
Crocker 1985, VandenBerg 1985, Chaboyer, Sarajedini \& Demarque 1992, Chieffi,
Straniero \& Salaris 1991, Salaris, Chieffi \& Straniero 1993, VandenBerg
1992, VandenBerg \& Bell 2001 and references therein) but to quantitatively
analyse the effect of the quoted very recent specific solar mixture variation
on Pop.II and Pop.I stellar cluster age determination.

To our knowledge all the extended sets of evolutionary tracks and isochrones
available in the literature (see e.g. Pietrinferni et al. 2004, Cariulo,
Degl'Innocenti, Castellani 2004, Castellani et al. 2003, VandenBerg et
al. 2001, Yi et al. 2001, Salasnich et al. 2000, Maeder \& Zahn, 1998 and
references therein) adopt a heavy element mixture older than Asplund et
al. 2004.

In this work we calculated evolutionary tracks and stellar isochrones for
 different mixtures with chemical compositions typical respectively of
 globular clusters (Z=0.0002, Z=0.001, Y=0.23) and open clusters (Z=0.02
 Y=0.27) in our Galaxy and of the bulk of the population of the Large
 Magellanic Clouds (Z=0.008 Y=0.25). The results for other chemical
 compositions can be extrapolated from the previous results. Calculations are
 presented in Sect.II while the theoretical results are discussed in Sect.III for
 globulars and in Sect.IV for intermediate age clusters.  In Sect. V
 theoretical predictions for the Hyades cluster are compared with
 observations.

\section{The Models and the physical inputs}

The models in this study have been computed with an updated version of the
FRANEC evolutionary code (see e.g. Chieffi \& Straniero 1989) adopting the
radiative opacity by the Livermore group (Iglesias \& Rogers 1996) and updated
nuclear cross sections (for more details see Cariulo, Degl'Innocenti,
Castellani 2004). Element diffusion has been included (Ciacio, Degl'Innocenti,
Ricci 1997, with diffusion coefficients from Thoul, Bahcall \& Loeb
1994). Radiative acceleration (see e.g. Richer et al. 1998, Richard et
al. 2002) has not been implemented. For convective mixing, we adopt the
Schwarzschild criterion to define regions in which convection elements are
accelerated (see the description in Brocato \& Castellani 1993).

With these choices we have already shown that theoretical predictions for the
color-magnitude diagram (CMD) of the nearby open clusters Hyades, Pleiades and
Ursa Major appear in good agreement with the observations for which precise
parallaxes are available from the Hipparcos satellite (Castellani,
Degl'Innocenti, Prada Moroni 2001, Castellani et al. 2002).

In the present models we adopted the EOS\_2001 by OPAL 
\footnote{http://www-phys.llnl.gov/Research/OPAL/Download/} and the conductive
opacity by Potekhin 1999, see also Potekhin et
al. 1999.

The change of the solar mixture affects age determination. For globular
clusters a widely used age indicator, independent of the cluster distance, is
the difference in visual magnitude between the Turn-Off and the Zero Age
Horizontal Branch at the RR Lyrae region ($\Delta$M$_V$(ZAHB-TO)).  For 
clusters up to about 6 Gyr, a useful age indicator is the difference in
visual magnitude between the He clump and the main sequence termination, MT,
($\Delta$M$_V$(clump-MT)). MT is evaluated at the maximum luminosity reached
just after the overall contraction (H exhaustion) whereas the clump magnitude
at the minimum luminosity of the He clump region.

In principle a mixture variation influences the evolution of a star on both
the burning (which is affected by the total CNO abundance) and the opacity, as
noted in studies of the effects of the enhancement of $\alpha$ elements in
globular cluster stars (see e.g. Simoda \& Iben 1970, VandenBerg 1985,
Chaboyer, Sarajedini \& Demarque 1992, Salaris, Chieffi, Straniero, 1993,
Salaris \& Weiss, 1998).  To account for both aspects we calculated the
opacities for the Asplund et al. (2004) mixture (URL:
http://www-phys.llnl.gov/Research/OPAL/new.html).

We point out that low temperature opacities are not available for the Asplund
et al. (2004) mixture; in particular the Alexander \& Ferguson (1994)
opacities, adopted in this work for T$\leq$12000 $^{\mathrm o}$K, are available only for
the Grevesse (1991) mixture. This is not a problem because it has already been
demonstrated by several authors (e.g. Rood 1981, Bazzano et al. 1982, Salaris
et al. 1993) that the main characteristics of population II stars are not
influenced by the low temperature opacities.  We also notice that model
atmospheres with the Asplund et al. (2004) solar composition are still not
available thus for our calculations we must adopt color transformations for
the old solar mixture (Castelli 1999, see also Castelli, Gratton \& Kurucz
1997). However preliminary calculations of stellar fluxes with the Asplund et
al. composition show a change of the stellar colors, with respect to the ones
calculated for the GN93 composition, within the
observational errors (Aufdenberg, private communication).

As firstly predicted by Simoda \& Iben (1968,1970) and Renzini
(1977) and confirmed by several authors (see e.g. Rood 1981, Bazzano et
al. 1982, Salaris et al. 1993) the TO characteristics and the mass of the He
core at the central helium ignition are mainly influenced by the burning and
thus by the CNO global abundances.  

The computed evolutionary models for intermediate age clusters cover with a
fine grid the mass range 0.7 to 8 M$_{\odot}$ for the adopted chemical
compositions Z=0.008 Y=0.250, Z=0.02 Y=0.27, where the amount of original
helium has been evaluated by assuming a primordial helium abundance Y$_P=0.23$
and $\Delta$Y/$\Delta$Z $\sim$2.5 (see e.g. Pagel \& Portinari 1998,
Castellani, Degl'Innocenti, Marconi 1999); the related isochrones have been
calculated from 100 Myr to 6 Gyr to properly cover the range of ages for the
Galactic open clusters. For globular cluster models we calculated masses from
0.6 to 1.0 M$_{\odot}$ for Z=0.0002 Y=0.23 and Z=0.001 Y=0.232 and isochrones
in the age range 8$\div$15 Gyr.  To select only the effects of the CNO
abundance on the H burning we calculated models of different masses (0.8
M$_{\odot}$, 1.2 M$_{\odot}$, 3.0 M$_{\odot}$, 5.0 M$_{\odot}$) for each of
the selected chemical compositions with the same physical inputs, included
radiative opacity table for the Grevesse \& Noels (1993) solar mixture, but
the C, N, O abundances of Asplund et al. (2004).  We found that in all cases
models with the new CNO abundance are quite identical with the old ones; the
variation of the total CNO abundance (of the order of 7\%) is perhaps too
small to be relevant for the burning.

\section{Results for globular clusters}

We analyzed isochrones for Z=0.0002 and Z=0.001 and ages from 8 to 15 Gyr.
The change of the heavy element mixture (both in the CNO abundance and in the
opacity calculations) does not affect neither the TO luminosity (see Fig. 1)
nor the ZAHB luminosity level. 

%It's well known (see e.g. Salaris et al. 1993)
%that, at fixed total metallicity, if the ratio of the abundance of elements
%with high first ionization potential and low first ionization potential is
%conserved
%($\frac{(X_C+X_N+X_O+X_{Ne})}{(X_{Mg}+X_{Si}+X_S+X_{Ca}+X_{Fe})}\approx$
%constant), models with a mixture variation are very similar. In our case,
%passing from Grevesse \& Noels (1993) to Asplund et al. (2004) composition
%this ratio changes by about 28\%, however the new models show the same TO and
%ZAHB luminosities as the old ones.  

It's worth clarifying a point: several observational works
demonstrate that in halo stars the $\alpha$ elements (e.g. O, Ne, Mg, Si, S,
Ca, Ti etc..) are enhanced by about the same amount with respect to iron in
comparison with the solar composition (see e.g. Gratton et al. 2003, Gratton
et al. 2000) and this may seem to be an additional problem in our analysis of
the dependence of the globular cluster age on the heavy elements mixture.
Luckily this is not the case; Salaris et al. (1993) suggested, for the first
time (see also e.g. Weiss, Peletier, Matteucci 1995, Salaris \& Weiss, 1998) ,
that for globular cluster stars, instead of calculating models with the
$\alpha$ enhanced mixture for Z$_0$, a good approximation can be obtained with
standard solar mixture isochrones of the metallicity Z${\mathrm tot}$ given by
Z${\mathrm tot}$=Z$_0$(af$_{\alpha}$ + b) with
f$_{\alpha}$=10$^{[\alpha/\mathrm{Fe}]}$.  The values of the coefficients $a$
and $b$ depend on the heavy-elements distribution; Salaris et al. 1993 give
$a$= 0.638 and $b$= 0.364, while with the Asplund et al. (2004) composition
they become respectively 0.659 and 0.341. Thus also for globular cluster stars
it's a righ procedure to restrict our analysis to the effects of the latest
solar mixture changes at fixed metallicity.

\section{Results for open clusters}

We adopt as age indicator the $\Delta$M$_V$(clump-MT).  The updating of the
heavy element mixture has, as for the globular cluster stars, no effect on the
MT luminosity and age.  Figure 2 shows the comparison of
$\Delta$M$_V$(clump-MT) as a function of age for Z=0.008 and Z=0.02 and the
labeled heavy mixtures.  The difference in the inferred age, which is thus
only due to the difference in the clump luminosity, increases with age and the
total metallicity reaching a maximum of the order of 0.5 Gyr in the present
range of ages. Altough not negligible, this discrepancy is still within the
uncertainty on the age determination with the adopted method, which can be
estimated of the order of $\approx$ 2 Gyr (see e.g. Cassisi et al. 1999,
Castellani \& Degl'Innocenti 1999). The change of the He burning evolutionary
times due to the mixture updating is negligible.

\section{Comparison with the Hyades}
As mentioned in the introduction, the variation in the solar heavy-element
mixture affects not only the theoretical tracks and isochrones for a given
global metallicity Z, but also the conversion of the spectroscopically
determined value of [Fe/H] to the total metallicity Z of the observed stars.
This is a quite tricky point because for computing 
stellar models one needs the total metallicity Z and 
not only the iron abundance, but to obtain the former from
the latter is necessary specifying the distribution of the element
abundances. Unless explicitely stated, it is usually assumed 
a solar mixture. Thus a revision of the 
photospheric abundance of the Sun directly implies a variation of the 
inferred total metallicity from the observed [Fe/H].
For solar-like stars, the [Fe/H] is often estimated by means of 
relative measures, that is by the comparison of the iron lines of the 
star and those of the Sun. In such a case, the numerical value
of [Fe/H] remains unaffected by a change of the solar mixture and the 
updated global metallicity of the star can be easily derived by adopting 
the new value for (Z/X)$_{\odot}$. 
In practice: 
$$
log\left(\frac{Z}{X}\right)_*= [Fe/H] +  log\left(\frac{Z}{X}\right)_{\odot}.
$$

As a consequence of the new value of (Z/X)$_{\odot}$=0.0165, the estimated
metallicity of the open clusters in the solar neighborhood has significantly
decreased. To analyse the effects of this change we need very precise
observational data.  To this aim we will focus our attention on the Hyades
cluster, for which very accurate Hipparcos observational data for the
distances (and thus for the absolute visual magnitude of the stars) are
available (see e.g. Dravins et al. 1997, Madsen, Dravins, \& Lindegren
2002). Moreover, for this cluster, there is negligible
 reddening (see e.g. Perryman et al. 1998).  Clusters with higher
uncertainties on the distance modulus and the reddening are not suitable,
in our opinion, for this analysis because the quoted uncertainties make
the discussion of the results much less clear.

We adopted for this cluster [Fe/H]= 0.14 $\pm$ 0.05 (see the discussion in
Perryman et al. 1998) in agreement with Paulson, Sneden \& Cochran 2003
too. Some years ago we already found a very good agreement with the Hyades
observations for a 520 Myr isochrone by adopting a metallicity of Z=0.024, a
value directly obtained from [Fe/H]= 0.14 and (Z/X)$_{\odot}$ by Grevesse \&
Noels 1993 (Castellani, Degl'Innocenti \& Prada Moroni 2001, Castellani et
al. 2002) in agreement with the results of other authors (see e.g. Perryman et
al. 1998, Lebreton 2000, de Bruijne, Hoogerwerf, \& de Zeeuw 2001, VendenBerg
\& Clem 2003). However, the updated (Z/X)$_{\odot}$,
reduces the global metallicity to Z=0.016. This quite large decrease of the
metallicity has a much greater effect on the fit of the cluster than the
change in the heavy element mixture adopted in the computation of the models.
In fact, by varying the solar mixture at fixed metallicity the MS color
remains almost the same with a maximum variation lower than two hundredth of
magnitude while the lowering of the metallicity from Z=0.024 to 0.016 blue
shifts the MS by about 0.05 mag.

Figure 3 shows the color-magnitude diagram of the Hyades with superimposed the
theoretical isochrones with Z=0.016 Y=0.27 and the labeled ages. As one can
easily see, the model now disagrees with data.  This result is
particularly relevant for the upper (B-V $\lcu$ 0.4) part of the MS, where, due
to the radiative envelope of the stars, the color is independent on the chosen
mixing length parameter (see e.g. Fig.2 of Castellani et al. 2001).  
Figure 3 also shows that changing the assumed age does not remove the disagreement.

The independence of the mixing length parameter also holds in the lower part (B-V
$\gcu$ 1.2) of the MS, where envelope convection is adiabatic; however in
this region the obtained results are less firm because, as well known, at low
temperatures a careful treatment of equation of state and molecular opacities
is needed.

A reasonable reduction of the helium abundance
doesn't solve the problem; to have an estimate of the shift of the MS position
as a function of the helium abundance see e.g. Castellani, Degl'Innocenti \&
Marconi (1999).

In order to reach a reasonable, although not perfect, match of the Hyades we
have to increase the metallicity at least to Z=0.02, corresponding to [Fe/H]=
0.23. 

Let us recall that we adopt color transformations and
bolometric corrections computed by adopting the Grevesse \& Noels 1993 
solar mixture. For a precise and self-consistent comparison one should 
use the same distribution of elements both for the interior and for the 
atmosphere, but, at present, the color transformations and the bolometric
corrections have not yet been updated. On the other hand, preliminary 
model atmosphere computations (Aufdenberg 2005, private communication) showed 
a minor effect.   

We also note that conclusions based on the MS position in the
color-magnitude diagram must be taken with caution because this is
 influenced at some level by the still present uncertainties on the physical inputs
adopted in the calculations and on the color transformations (see
e.g. Castellani et al. 2001, Sekiguchi \& Fukugita 2000).

\section*{Acknowledgments}
We are extremely grateful to V. Castellani G. Fiorentini and S. Shore for
useful discussions and for a careful reading of the paper.  We warmly thank
J. Aufdenberg for his preliminary analysis of the stellar fluxes with the
Asplund et al. 2004 composition. Financial support for this work was provided
by the Ministero dell'Istruzione, dell'Universit\`a e della Ricerca (MIUR)
under the scientific project ``Continuity and discontinuity in the Galaxy
formation'' (P.I.: R. Gratton).  \\

%==============================================  FIGURA  1
\begin{figure}
\label{TO}
\centerline{\epsfxsize= 10 cm \epsfbox{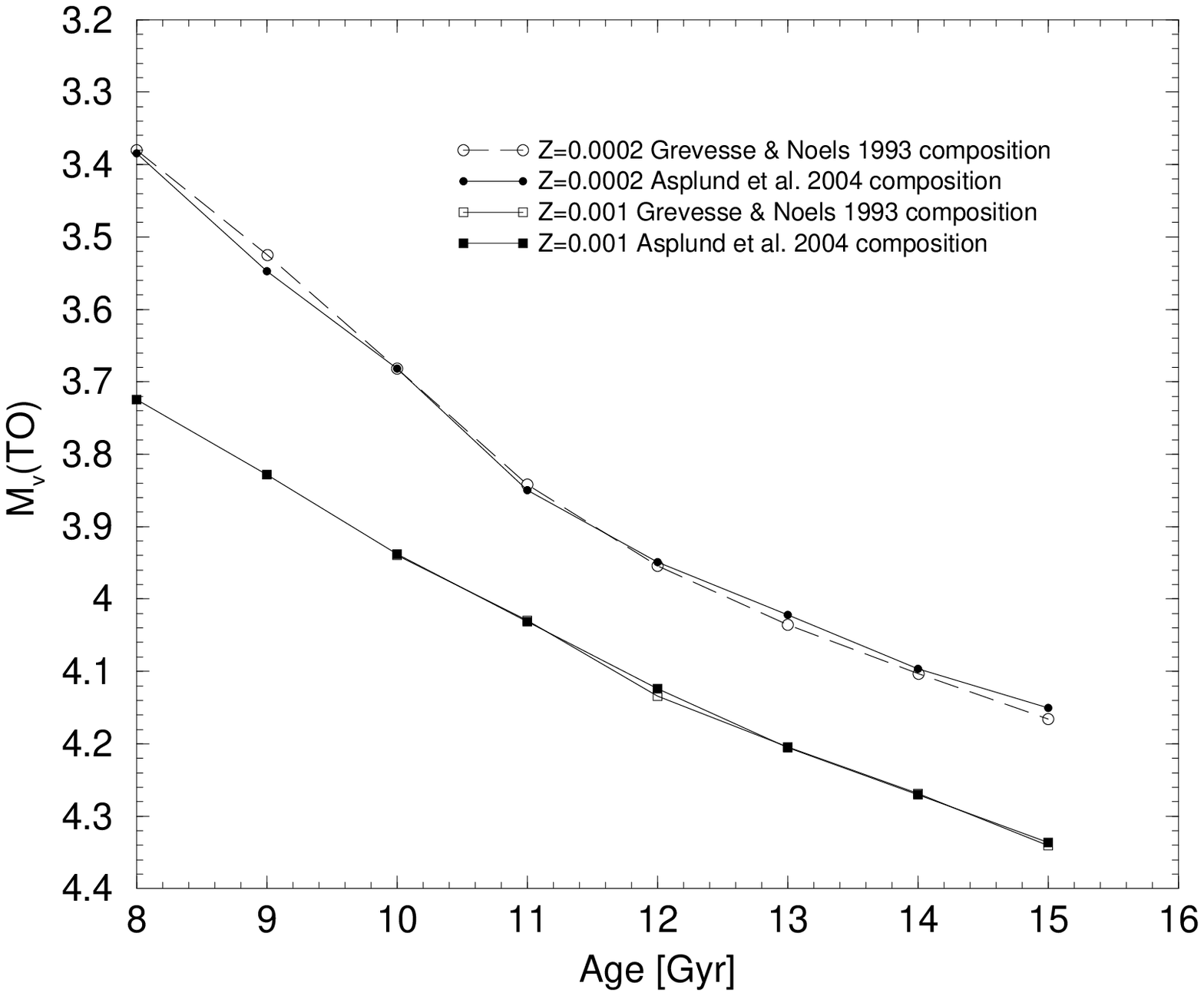}}
\caption{The TO absolute visual magnitude, M$_V$(TO), as a function of
the cluster age for the labeled assumptions about the original
chemical composition and the heavy element mixture.}
\end{figure}
%==============================================
\newpage
%==============================================  FIGURA  2
\begin{figure}
\label{MTclump}
\centerline{\epsfxsize= 10 cm \epsfbox{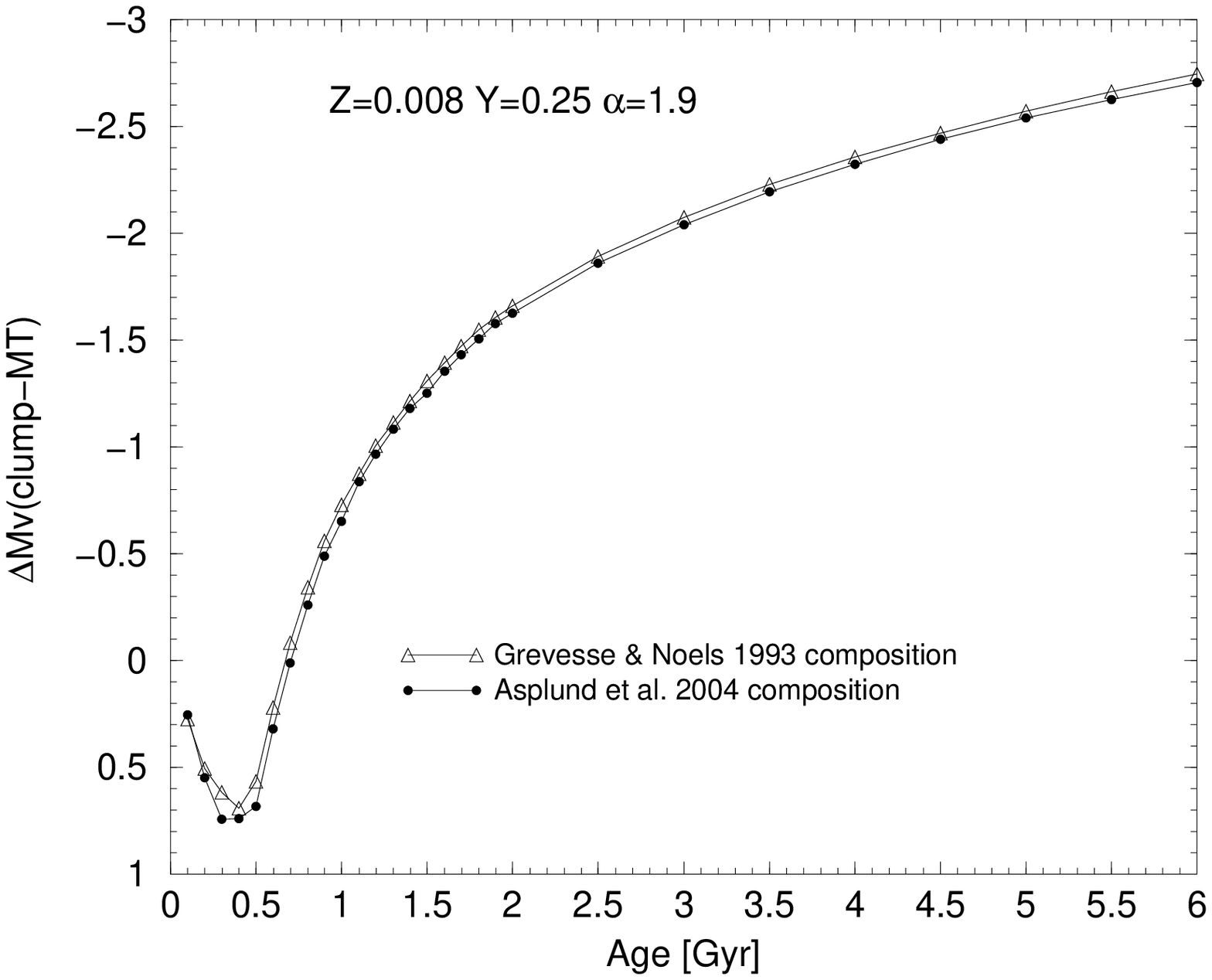}}
\centerline{\epsfxsize= 10 cm \epsfbox{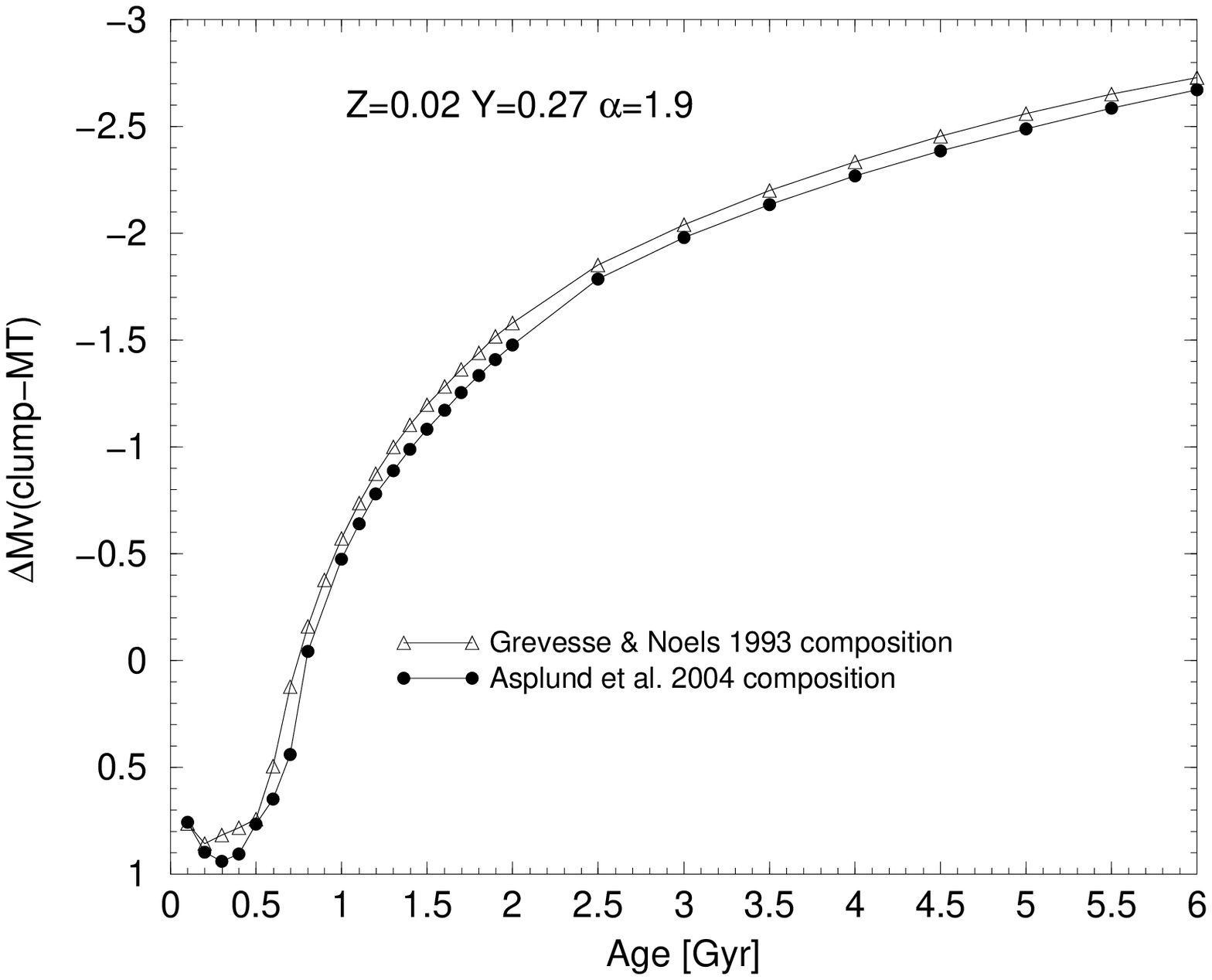}}
\caption{The $\Delta$M$_V$(clump-MT) as a function of
the cluster age for the labeled assumptions about the original
chemical composition and the heavy element mixture.}
\end{figure}
%================================================
\newpage
%==============================================  FIGURA  3
\begin{figure}
\label{IadiZ16.eps}
\centerline{\epsfxsize= 10 cm \epsfbox{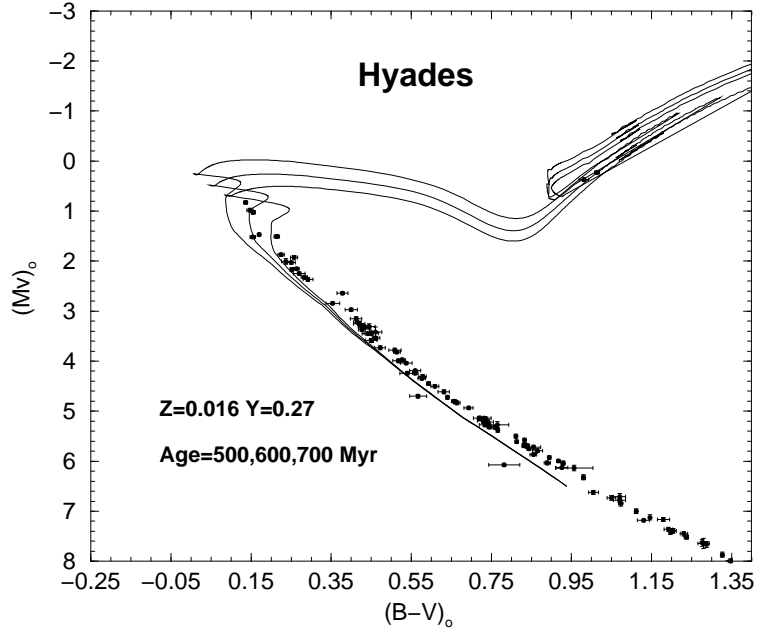}}
\caption{The CMD for the Hyades, using the parallax
values from Madsen et al. (2002). Visual, spectroscopical and
suspected binaries are excluded, see also Madsen et al. (2000).
Error bars indicate observational errors as
given by Madsen et al. (2002) for the parallax and by the Hipparcos
catalog (at the node http://astro.estec.esa.nl/Hipparcos/HIPcataloguesearch.html) for the
colors. Observational data are compared with present theoretical isochrones 
for Z=0.016 Y=0.27 $\alpha$=1.9. Color transformations and
bolometric corrections from Castelli (1999).}
\end{figure}
%================================================

\end{document}